# Specific Heat of Kagomé Ice in the Pyrochlore Oxide $Dy_2Ti_2O_7$

Zenji Hiroi, Kazuyuki Matsuhira[1], Seishi Takagi[1], Takashi Tayama and Toshio Sakakibara

*Institute for Solid State Physics, University of Tokyo, Kashiwa, Chiba 277-8581*
[1]*Department of Electronics, Faculty of Engineering, Kyushu Institute of Technology,
Kitakyushu 804-8550*

A novel macroscopically degenerate state called kagomé ice, which was recently found in a spin ice compound $Dy_2Ti_2O_7$ in a magnetic field applied along the [111] direction of the cubic unit cell, is studied by specific heat measurements. The residual entropy of the kagomé ice is estimated to be 0.65 J/K mol Dy, which is nearly 40 % of that for the tetrahedral spin ice obtained in a zero field (1.68 J/K mol Dy) and is in good agreement with a theoretical prediction. It is also reported that the kagomé ice state, which is stabilized at a range of magnetic field of 0.3 ~ 0.6 T, is a "gas" phase and condenses into a "liquid" phase with nearly zero entropy at a critical field of 1 T.

## §1. Introduction

Geometrical frustration is one of the most interesting topics in studying magnetism of solids. The pyrochlore lattice made of corner-sharing tetrahedra is a typical three-dimensional network exhibiting the geometrical frustration and has been studied extensively. In the course of searching novel materials related, intriguing compounds were found in pyrochlore oxides comprising the pyrochlore lattice, which is $Dy_2Ti_2O_7$ and $Ho_2Ti_2O_7$.[1,2] In these compounds an Ising spin with a large magnetic moment of $10\mu_B$ from the rare-earth ion resides on the pyrochlore lattice and weakly interacting ferromagnetically with its neighbors. Moreover, the local quantization axis is fixed to the <111> directions of the cubic unit cell due to the crystal field effect. As a consequence, a two-in, two-out spin configuration in each tetrahedron is preferred, which is called the ice rule, and a macroscopic degeneracy in the ground state analogous to water ice is realized.[3] Apparently, this spin ice state provides us with a new strategy for the research of frustrated magnets.

Recently, Matsuhira et al.[4] and Higashinaka et al.[5] reported another macroscopically degenerate state in $Dy_2Ti_2O_7$ in a magnetic field applied along the [111] direction. The pyrochlore lattice consists of triangular and kagomé planes stacked alternatingly along the [111] direction, as shown in Fig. 1a. Since the spins on the former planes possess their easy axis along the field direction, they are pinned easily along the field to gain Zeeman energy. Even in this situation, however, the ice rule stabilizing the two-in, two-out configuration can be satisfied in a low magnetic field, as depicted in Fig. 1b. In other words, applying the field selects some of two-in, two-out configurations with the apical spin fixed along the field out of macroscopically degenerate spin ice states. Now we have two kinds of tetrahedra, A and B, with an apical spin above and below the basal triangle in the kagomé plane, respectively. In tetrahedron A the three kagomé spins should have an in-in-out (iio) configuration, because the apical spin points already outward, while in B an in-out-out (ioo) configuration should be realized. As a result, a modified ice rule (iio-ioo rule) is to be applied to spin configurations in the kagomé plane. Obviously, a reduced macroscopic degeneracy should remain for the kagomé lattice. This means that the spin dimensionality is controlled successfully by the magnetic field. Matsuhira et al. called this new state "kagomé ice", and reported its residual entropy of about 0.8 J/K mol Dy.[4] Stimulated with this finding, a theoretical calculation was carried out by Udagawa et al.[6] They obtained the exact value for the ground-state entropy to be $0.0808R$ (0.67 J/K mol Dy) where $R$ is the gas constant.

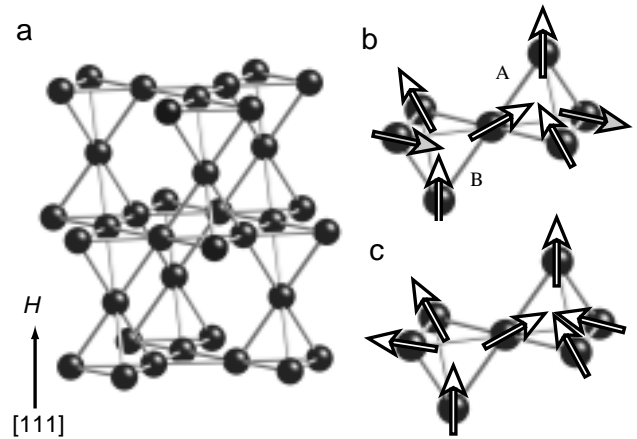

Fig. 1. Pyrochlore lattice (a) consisting of corner-sharing tetrahedra. Balls represent Dy atoms in $Dy_2Ti_2O_7$. The structure contains triangular and kagomé planes alternately stacked along the [111] direction. Local spin arrangements on a couple of tetrahedra, which are stabilized at low and high fields along the [111] direction, are shown in (b) and (c), respectively. Arrows show the orientation of Ising spins on $Dy^{3+}$ ions. The shaded and open arrows indicate spins with and without an antiparallel component to the field, respectively. Tetrahedra A and B with apical spins outward and inward, respectively, should be distinguished from each other under the magnetic field along [111].

The difference between the experimental and calculated values may come from experimental difficulty in determining residual entropy, the detail of which we will discuss later.

More recently, Sakakibara *et al.* found an unique liquid-gas (*l-g*) phase transition for the present system in magnetic fields along the [111] direction.[7] They measured magnetization down to 50 mK and observed a jump at a critical field $H_c$ near 0.9 T below $T = 0.36$ K. Below $H_c$ a plateau-like structure with a magnetic moment close to 3.3 $\mu_B$/Dy is seen, which evidently corresponds to the kagomé ice state, while another plateau above $H_c$ with a magnetic moment of ~ 5 $\mu_B$/Dy corresponds to the fully saturated state shown in Fig. 1c.

Here we report the specific heat of $Dy_2Ti_2O_7$ down to 0.4 K in various magnetic fields. The field dependence of residual entropy was obtained. We will discuss the origin of the phase transition and suggest that the kagomé ice state is a "gas" phase and the fully saturated state is a "liquid" phase.

## §2. Experimental

A single crystal of $Dy_2Ti_2O_7$ was prepared by the floating-zone method in an infrared furnace with four halogen lamps and elliptical mirrors (Crystal Systems, Inc.). It was grown under oxygen gas flow with a typical growth rate of 4 mm/h. The obtained single crystal was translucent yellow. A few small, thin plate-like crystals were cut from the boule for specific heat measurements. The typical size and weight were $1 \times 1 \times 0.1$ mm$^3$ and 1 mg. To minimize a demagnetizing field effect, the plate surface was chosen to be (011), so that a magnetic field along [111] or [100] was parallel to the plate surface. A correction of applied magnetic field due to that effect was not taken into account in the present study.

Specific heat was measured by the heat-relaxation method in a temperature range between 0.37 K and 70 K in a Quantum Design Physical-Property-Measurement-System. A single crystalline sample was attached to a sapphire substrate by a small amount of Apiezon N grease. An addenda heat capacity had been measured in a separate run without the sample, and was subtracted from the data. A magnetic field was applied approximately along the [111] or [100] direction of the cubic unit cell. The accuracy of the orientation may be within a few degrees. We carried out experiments several times in the same way and confirmed that the effect of misalignment of samples was negligible except for conditions of $H \sim 1$ T and $T < 0.6$ K.

## §3. Results

### 3.1 Specific heat at zero field

Figure 2 shows the temperature dependence of specific heat $C$ per one mol of Dy atoms at zero field. It exhibits a broad peak around 1 K without any signs for long-range order down to 0.4 K, in good agreement with previous results.[3] The specific heat of insulating $Dy_2Ti_2O_7$ consists of a magnetic contribution $C_m$ and a lattice contribution $C_l$. The latter is expressed to be $\alpha T^3$ in the low-temperature limit. This lattice contribution is clearly seen in a $C/T$ versus $T^2$ plot shown in the inset to Fig. 2, where a linear variation is seen for 12 K $\leq T \leq$ 19 K. It means that the magnetic contribution is negligible above $T = 12$ K. The $\alpha$ value is determined to be $4.85 \times 10^{-4}$ JK$^{-4}$/mol Dy, which gives a Debye temperature of 353 K, a typical value for pyrochlore oxides.[8]

Figure 3 shows the temperature dependence of $C_m/T$ obtained for two crystals with different weights of 1.57 mg and 0.548 mg by subtracting the above-mentioned lattice contribution. The two sets of data exactly trace the same curve, indicating a high quality of the data. The magnetic entropy $S(T)$ is estimated by integrating $C_m/T$ from 0.4 K to $T$;

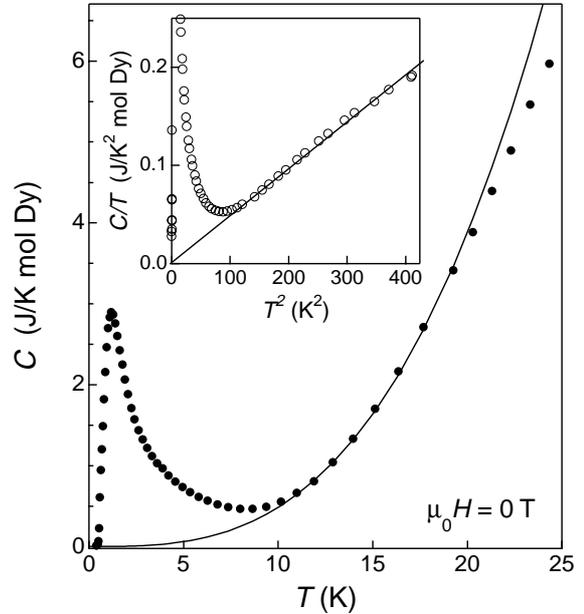

Fig. 2. Specific heat *C* measured at zero field. The solid line shows a calculated lattice contribution, $C_l = \alpha T^3$ with $\alpha = 4.85 \times 10^{-4}$ JK$^{-4}$mol$^{-1}$, which is determined from the $C/T$ versus $T^2$ plot shown in the inset.

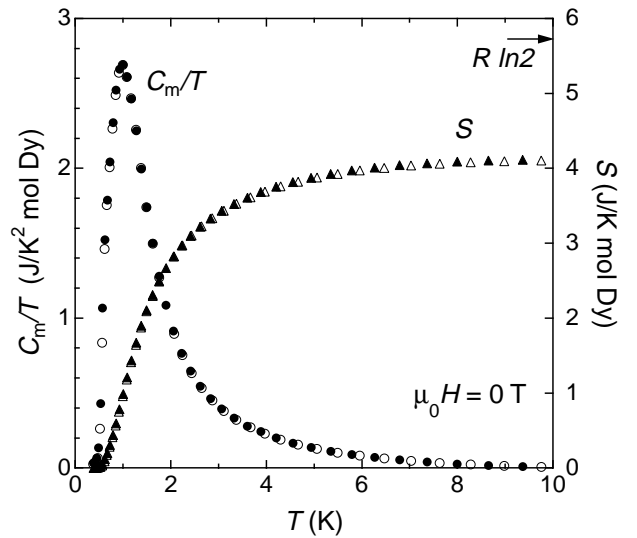

Fig. 3. Magnetic specific heat divided by temperature $C_m/T$ on the left axis and entropy *S* on the right axis. Open and solid marks represent data measured on two crystals with different weights of 1.57 mg and 0.548 mg. The saturation value of *S* at high temperature is smaller by 1.66 JK$^{-1}$mol$^{-1}$ than the expected total entropy *R*ln2, which corresponds to the zero-point entropy of the spin ice.

$$S(T) = \int_{0.4}^{T} \frac{C_m}{T} dT$$

and is plotted in Fig. 3. The temperature dependence is almost saturated at 10 K with a value of 4.10 JK$^{-1}$mol$^{-1}$, which is apparently much smaller than $R\ln 2$ (5.76 JK$^{-1}$mol$^{-1}$) thermodynamically expected for an Ising spin system with two energy levels. The previous, similar observation on a polycrystalline sample leads Ramirez et al. to conclude the spin ice ground state with residual entropy.[3] Our $C_m/T$ data is slightly different from theirs: the peak height near 1 K is 2.67 JK$^{-2}$mol$^{-1}$ in our data which is considerably larger than theirs (about 2.4 JK$^{-2}$mol$^{-1}$). This difference may give rise to a larger entropy value in our case. We estimate the residual entropy for the spin ice to be 5.76 - 4.10 = 1.66 JK$^{-1}$mol$^{-1}$, which is very close to the Pauling's prediction, 1.68 JK$^{-1}$mol$^{-1}$. It suggests that the above estimation of the lattice part is reasonable.

*3.2 Field dependence: H // [100]*

First we describe the change of specific heat under a magnetic field parallel to the [100] direction. As already reported by Monte Carlo simulations and magnetization measurements, the destruction of the spin ice state to a fully saturated state occurs at low fields along the [100] direction with no intermediate states.[9,10] This is because the magnetic field along [100] selects only one spin configuration as depicted in the inset to Fig. 4, without breaking the ice rule. Although the internal energy from spin interactions may not be affected, the gain of Zeeman energy compensates the loss of entropy.

We carried out specific heat measurements up to $H = 2.0$ T and evaluated the change of entropy. Measurements were done up to 30 K for $H \leq 1$ T and 50 K for $H = 2.0$ T. It was necessary to measure specific heat up to higher temperature with increasing field, because entropy was released at higher temperature. Experimental determination of a magnetic contribution becomes more difficult at higher temperature, mostly because of rapidly increasing lattice contributions. Another contribution from the first excited state for a Dy$^{3+}$ ion, which exists about 150 K above the ground state,[11] may also cause an experimental error. We used the lattice contribution below $T = 12$ K as determined by the zero-field experiment. On the other hand, a magnetic contribution above 12 K was obtained simply by subtracting a zero-field data from nonzero-field data, although this might include certain ambiguities.

As shown in Fig. 4a, the peak position in $C_m/T$ shifts to higher temperature with increasing the magnitude of field. Accordingly, the entropy curve shifts to higher temperature as shown in Fig. 4b. Here a certain value is added to each set of experimental values, so that the saturated value at the highest temperature becomes $R\ln 2$. Since it is convenient to see how residual entropy $S_r$ is released with field, we use this modified $S$ in the following part of the paper. Note that the $S_r$ is removed almost completely at $H = 0.5$ T and then slightly increases with field. Especially in the $H = 2$ T data, the saturation at high temperature seems incomplete, suggesting the underestimation of magnetic entropy. Since the high-field state should have zero entropy, the observed increase in $S_r$ at high field must be an artifact. We will be back to this point later.

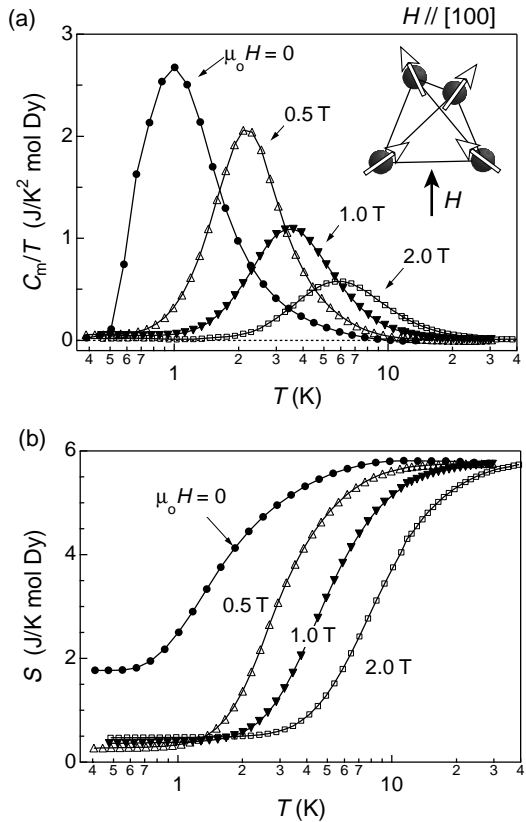

Fig. 4  Magnetic specific heat (a) and the entropy (b) at various magnetic fields applied along the [100] direction. Each set of entropy data has been added by a certain offset so as to make the highest-temperature value equal to $R\ln 2$.

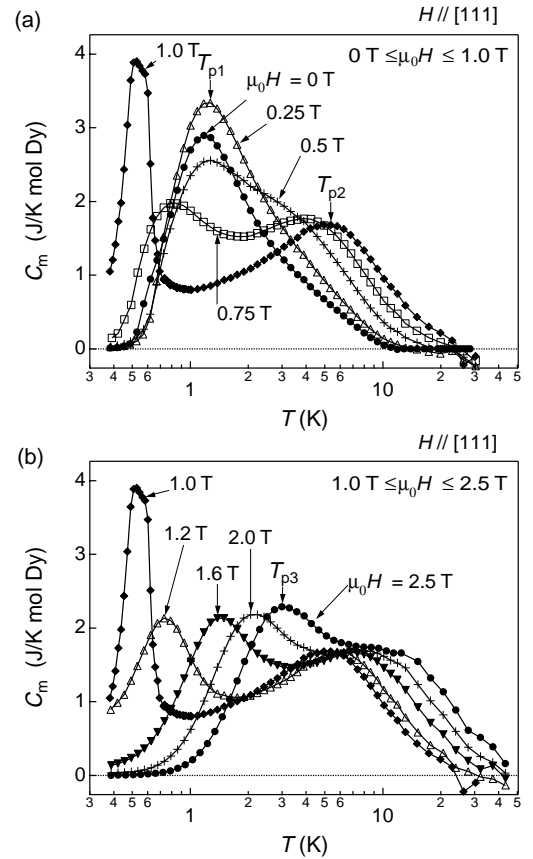

Fig. 5.  Specific heat measured at various magnetic fields parallel to the [111] direction. Low and high field data are shown separately in (a) and (b), respectively. The positions of three characteristic peaks, $T_{p1}$, $T_{p2}$, and $T_{p3}$ are marked with arrows.

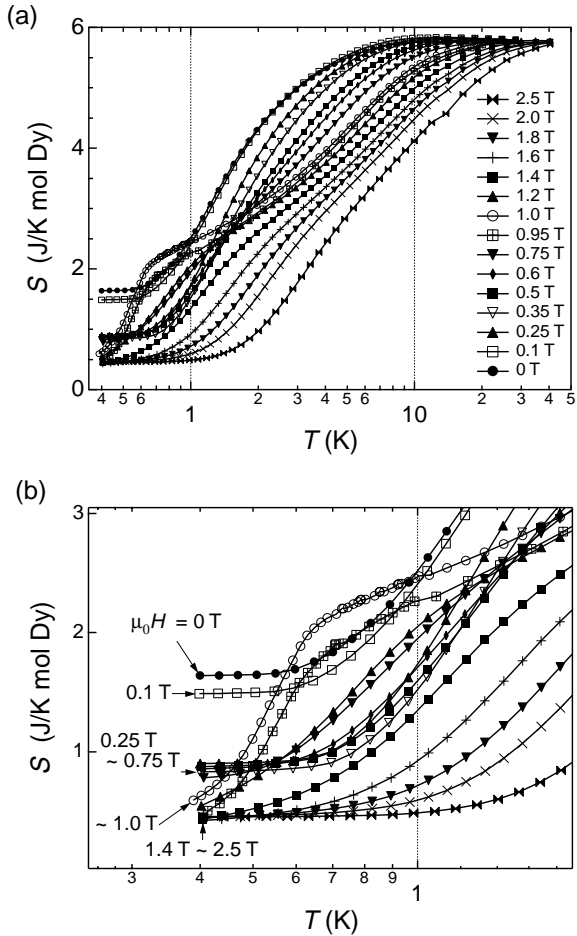

Fig. 6   Temperature dependence of entropy at various magnetic fields parallel to the [111] direction. The $S$ value at $T = 0.4$ K in each field except $H \sim 1$ T gives a good estimate for a residual entropy, because the temperature dependence has been already saturated there.

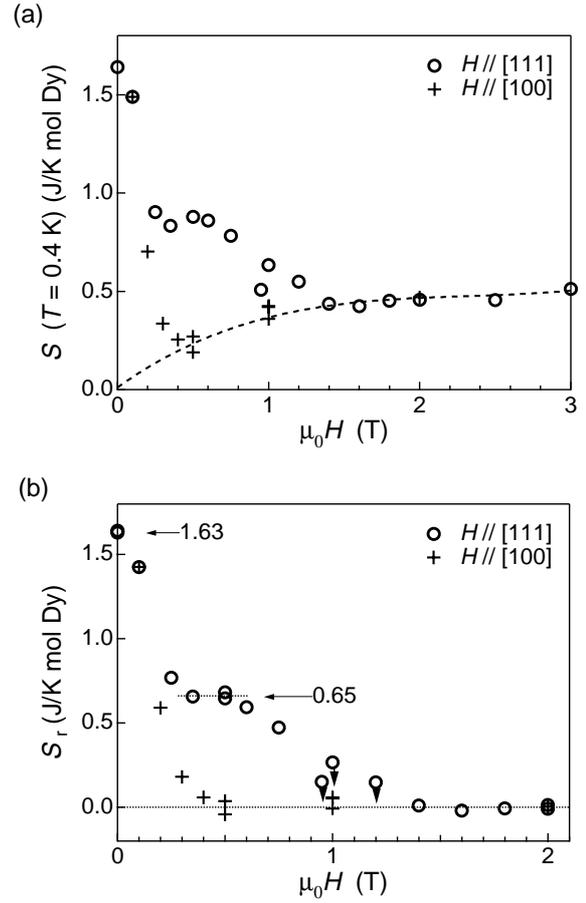

Fig. 7   Field dependence of entropy at $T = 0.4$ K obtained for two field directions (a) and corrected residual entropy $S_r$ (b). The dotted line in (a) is a fitting curve to the data of $H \; // \; [100]$ at $H \geq 0.5$ T, which is used as a background coming from experimental errors in determining $S$. The plateau in $S_r$ seen at $H \sim 0.5$ T for $H \; // \; [111]$ corresponds to the zero-point entropy of the kagomé ice. The $S_r$ data for $H \; // \; [111]$ near 1 T overestimates actual values because of proximity to the phase transition, and must be close to zero.

*3.3 Field dependence: $H \; // \; [111]$*

As reported in our previous study,[4] the field dependence of specific heat for $H \; // \; [111]$ is rather complicated, compared with the case for $H \; // \; [100]$. Figure 5 shows selected sets of data for fields below and above 1 T separately. With fields less than 0.25 T the peak in $C_m$ just grows upward with keeping the peak temperature ($T_{p1}$) nearly constant. Then, a new shoulder appears around $T = 3$ K at $H = 0.5$ T, which shifts to a high temperature side with field (Its peak temperature is called $T_{p2}$). Simultaneously, the first peak starts to shift to low temperature above $H = 0.5$ T. Surprisingly observed around 1 T is a sudden appearance of a sharp and intense peak below 0.6 K. Above 1 T another broad peak is observed at $T_{p3}$. Both $T_{p2}$ and $T_{p3}$ increase gradually with field, and no more drastic changes occur at higher fields up to 3.5 T. The origin for these peaks will be discussed later.

Figure 6 shows the temperature dependence of calculated entropy. The entropy curves gradually shift to high temperature with increasing field, as in the case for $H \; // \; [100]$, except for those near $H = 1$ T. However, the field dependence of residual entropy is quite different, depending on the field directions. For $H \; // \; [111]$ the $S_r$ does not decrease monotonously, but stays at a finite value near 0.8 JK⁻¹mol⁻¹ for $0.25$ T $< H < 0.75$ T, which gives a plateau-like change in the field dependence shown in Fig. 7. Near a critical field of 1 T, the decrease in $S$ has not yet saturated even near 0.4 K, which means that the residual entropy is lower than the value at 0.4 K. In contrast, the $S$ above the filed is almost saturated at 0.4 K to a small finite value, similar as in the case of $H \; //$

[100].

The field dependence of $S$ at $T = 0.4$ K with magnetic field along the two directions is summarized in Fig. 7a. It is obviously seen that only the data for $H$ // [111] exhibits a plateau-like variation. The two data sets seem to agree to each other at high fields, slightly increasing with field. This must be due to experimental difficulty in determining a magnetic specific heat at high temperature, as mentioned before. Therefore, this extrinsic background was estimated by fitting the data for $H$ //[100] at $H > 0.5$ T to a quadratic form, assuming that the $S_r$ is zero there. Figure 7b shows the field dependence of thus corrected $S_r$ after subtraction of the background. The $S_r$ of the intermediate state for $H$ // [111] is now determined to be 0.65 JK$^{-1}$mol$^{-1}$, which must be the residual entropy of the kagomé ice. The $S_r$ near $H = 1$ T must be nearly zero, because the $S$ still decreases at $T = 0.4$ K, as shown in Fig. 6b. Therefore, the $S_r$ for $H$ // [111] is released in two steps finally to the zero-entropy state above $H = 1$ T.

### 3.4 Phase transition

Figure 8 shows the $C_m/T$ near the critical field $H_c = 1$ T, where an intense peak around 0.5 K is seen only for fields of 0.95 T and 1.0 T. It collapses at 0.9 T and 1.1 T, followed by a broad peak at $H \leq 0.75$ T and $H \geq 1.2$ T, respectively. This indicates that a phase transition occurs at $H_c$ below $T_c \sim 0.5$ K. The strange shape of the peak suggests that the transition is accompanied by latent heat, indicative of the first-order transition. Unfortunately our relaxation method could not measure latent heat. We also measured specific heat in an isothermal condition at 0.4 K with varying fields. The result shows an anomaly around 1 T as in the inset to Fig. 8.

Figure 9 shows a series of isothermal entropy curves as a function of field obtained from the data sets measured as a function of temperature under various fields. There is a peak in $S$ at $H = 1$ T for $0.5 \leq T \leq 1.0$ K. This means that a large entropy remains near $H_c$ down to low temperature above $T_c$, and is released suddenly below $T_c$.

### 3.5 Phase diagram

All the specific heat data for $H$ // [111] is summarized in Fig. 10 in the form of magnetic field-temperature phase diagram, where the position of the three observed peaks is plotted. $T_{p1}$ which corresponds to the original peak at zero field is almost independent of field below $H = 0.5$ T, but suddenly decreases above that field toward the critical point at $H_c = 1$ T and $T_c \sim 0.5$ K. $T_{p2}$ becomes discernible above 0.5 T and gradually increases at high field. On the other hand, $T_{p3}$ appears above $H_c$ and increases gradually with field.

The two field-sensitive peaks at $T_{p2}$ and $T_{p3}$ should come from the Schottky anomaly of Zeeman-split spins. As shown in Fig. 1, the apical spin which belongs to the triangular plane of the pyrochlore lattice must be easily pinned along the field, because it possesses larger Zeeman energy than the other three spins on the kagomé plane. The Zeeman splitting $\Delta$ for a spin with a total angular moment $J$ at magnetic field $H$ is given by $2g_J J\mu_B H$, where $g_J$ is the Landé $g$-factor. The peak of a Schottky-type specific heat for such a two-level state is seen at $k_B T_p = 0.38\Delta$.[12] Thus, for a free Dy$^{3+}$ ion with $g_J J = 10$, one expects a peak at $T_p = 7.6(\mu_B/k_B)H = 5.10H$. This relation is plotted in Fig. 10, with which $T_{p2}$ agrees well at high field. The small deviation at low field must be due to weak interactions between neighboring spins. Therefore, it is concluded that the high temperature peak at $T_{p2}$ corresponds to the freezing of the apical spins. This means that only spins on the kagomé plane have degree of freedom below $T_{p2}$.

On the other hand, in the two-in, two-out state selected by the field, as shown in Fig. 1b, one of the three spins on each triangle in the kagomé plane should have a spin component antiparallel to the field. A Schottky peak for this spin is expected at $T_p = 5.10H/3$, which is also plotted in Fig. 10. $T_{p3}$ approaches asymptotically to the calculated curve at high field, suggesting that $T_{p3}$ is related to the freezing of this spin in the kagomé plane. Therefore, the fully saturated spin configuration shown in Fig. 1c is attained below $T_{p3}$. However, it is to be noted that there is no long-range order. The large deviation, compared with the case for $T_{p2}$, implies that correlations between spins in the kagomé plane are much

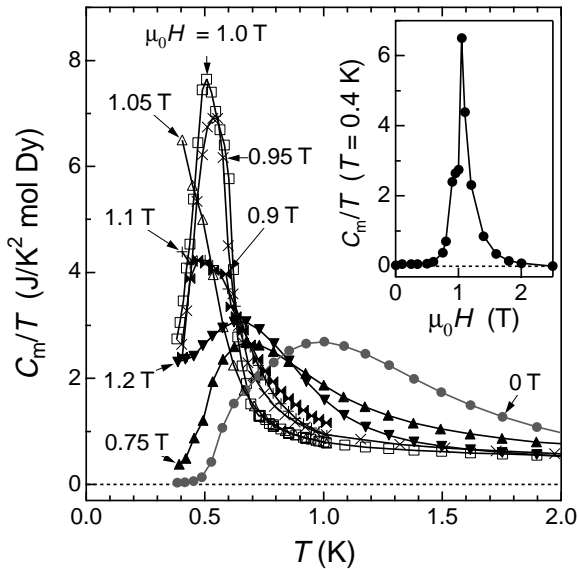

Fig. 8.  Specific heat for $H$ // [111] near $H = 1$ T showing a large peak around $T = 0.5$ K. The field dependence at $T = 0.4$ K is shown in the inset.

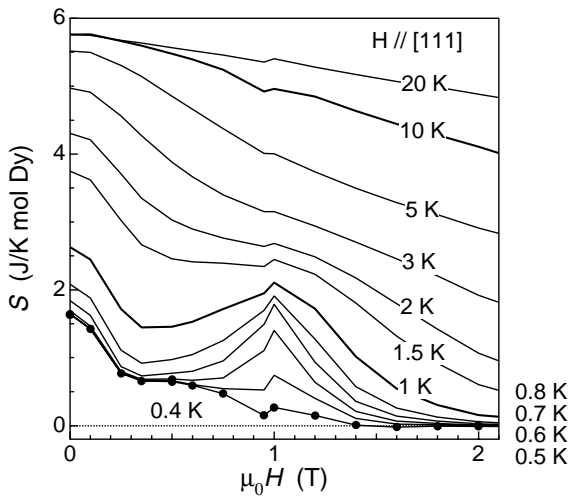

Fig. 9.  Isothermal entropy as a function of field at $T = 0.4 - 20$ K.

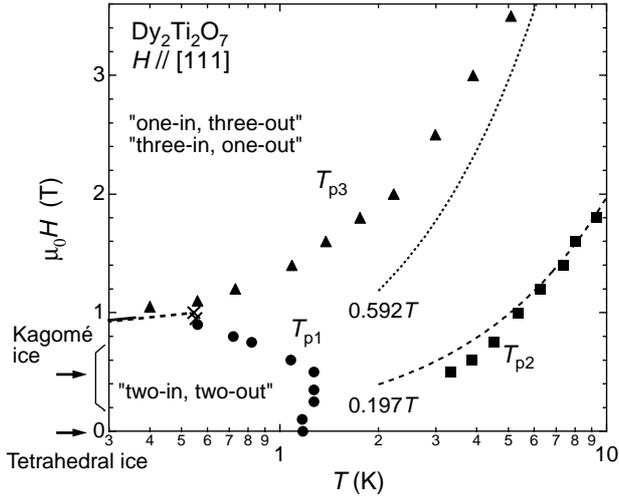

Fig. 10. Magnetic field-temperature phase diagram for $H \parallel [111]$. Circles, squares, and triangles represent the three peak positions found in the specific heat measurements, $T_{p1}$, $T_{p2}$, and $T_{p3}$, respectively. A critical point at $(T_c, H_c) = (0.5\,K, 1\,T)$ is shown by cross, and the associated first-order phase transition line by a thick dotted line. The phase transition line with a critical point $(T_c, H_c) = (0.36\,K, 0.93\,T)$ determined by Sakakibara et al.[7] is also shown by a thick line.

larger.

The peak at $T_{p1}$ must be distinguished from the other two peaks, because it is rather insensitive to field or even shift to low temperature with increasing field. It should arise from a certain short-range correlation in the kagomé plane, that is, freezing into the spin ice or kagomé ice state.

The critical point at $(T_c, H_c) = (0.5\,K, 1\,T)$ for the phase transition is shown by a cross in Fig. 10. It must correspond to that found by recent magnetization measurements: $(T_c, H_c) = (0.36\,K, 0.93\,T)$.[7] The small difference in $H_c$ may come from corrections due to the demagnetizing field effect, while the origin of the large difference in $T_c$ is not known. Possibly the $T_c$ is sensitive to a small misorientation of a crystal with respect to the field.

## §4. Discussion

### 4.1 Kagomé ice

We have shown that a new macroscopically degenerate state called kagomé ice is stabilized for magnetic fields along the [111] direction between 0.3 ~ 0.6 T. The residual entropy is found to be 0.65 JK$^{-1}$mol$^{-1}$, which is about 40 % of the tetrahedral spin ice. Here we discuss the origin of the residual entropy. As we have described in the introduction, now we have a kagomé plane with Ising spins where the modified ice rule should be applied. That is, the in-in-out configuration is satisfied for triangle A, while the in-out-out one for triangle B, as depicted in Fig. 11a. Spanning the whole kagomé lattice with spins in this way, a macroscopic number of ways are possible, one of which is illustrated in Fig. 11b, where only spins with an antiparallel component to the field are shown with arrows. Moving one spin to the other corners of the triangle should give rise to a local rearrangement of spin configurations by the domino effect, with keeping the modified ice rule. Any resulting state should have exactly the same energy as before. Therefore, there is a macroscopic degeneracy in the kagomé ice ground state, similar to the original tetrahedral spin ice.

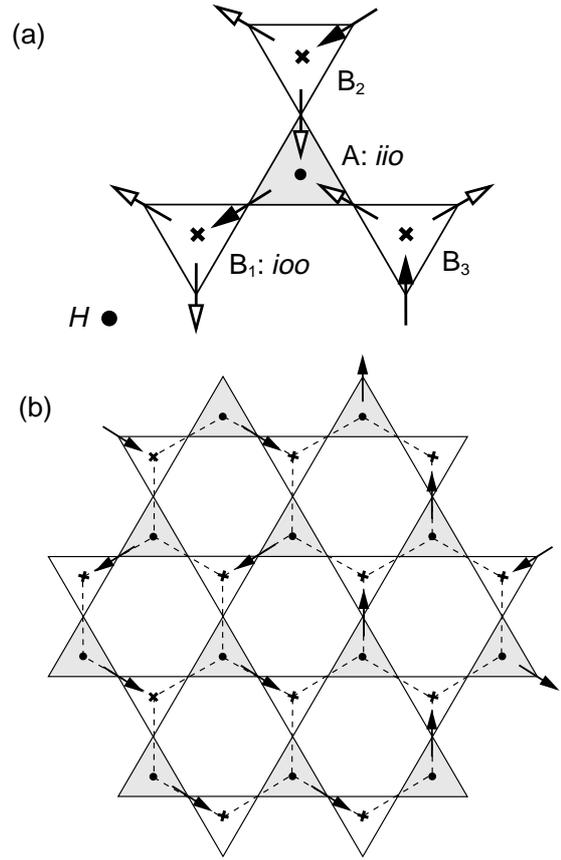

Fig. 11. (a) Schematic representation of a local spin arrangement showing the modified ice rule on the kagomé net. Black and white arrows indicate spins with and without an antiparallel component to the field, respectively. To keep the ice rule for a tetrahedron, in-in-out and in-out-out spin configurations are to be satisfied for triangles A and B, respectively, in the case of the kagomé lattice. When we fix such a spin arrangement for triangle A as illustrated in (a), still we have two ways of arranging spins for each triangle $B_2$ and $B_3$, while only one way for triangle $B_1$. Therefore, we expect $3 \times 2^2$ ways for this structural unit.[6] (b) One of global spin configurations on the kagomé net chosen from the macroscopically degenerate kagomé ice state. Moving one spin to the other corners of the triangle should give rise to a local rearrangement of spin configurations, with keeping the modified ice rule, by the domino effect. The resulting states should have exactly the same energy as before.

According to Udagawa et al., the residual entropy is calculated by the Pauling's method as follows:[6] Let's assume $N$ triangles of A, each of which is surrounded by three triangles of B. Since there are three ways of arranging spins for each triangle A, the number of ways for whole system, assuming a random spin arrangement, is $3^N$. Note that we do not have to think about triangle B explicitly, and that triangle A is connected to each other by the modified ice rule through triangle B. Next we will consider a unit composed of one triangle A and three triangles B shown in Fig. 11a, where there are $3^3$ ways of arranging spins without constraints. Among them only $3 \times 2 \times 2$ ways are allowed by the iio-ioo rule, because two ways of arrangement remain for triangles $B_2$ and $B_3$, while only one for $B_1$. Therefore, the total entropy is calculated by

$$S = k_B \ln\left(3^N \left(\frac{3 \times 2^2}{3^3}\right)^N\right) = Nk_B \ln\left(\frac{4}{3}\right).$$

As we have $4N$ spins in total, the $S_r$ per mol Dy is $(R/4)\ln(4/3) = 0.598$ JK$^{-1}$mol$^{-1}$. This value is slightly smaller than our experimental value.

On the other hand, Udagawa et al. found another solution by considering that this problem can be mapped onto a dimer covering problem on a honeycomb lattice which is exactly solvable.[6] As shown in Fig. 11b, a honeycomb lattice emerges from the kagomé lattice by taking the center of each triangle as a new lattice point. Then, putting spins with an antiparallel component to the field on the kagomé lattice under the modified ice rule is exactly the same as creating and dispersing dimers on the honeycomb lattice. They calculated the exact value for the $S_r$ to be 0.672 JK$^{-1}$mol$^{-1}$, which is very close to our experimental value of 0.65 JK$^{-1}$mol$^{-1}$.

*4.2 Liquid-gas phase transition*

The phase transition observed at $H_c \sim 1$ T is quite interesting. Obviously it occurs between the kagomé ice state and the fully saturated state. Since there is no symmetry breaking between the two phases, it must be a liquid-gas (*l-g*) type phase transition.[7] A similar *l-g* transition has been predicted by using Monte Carlo simulations for the present system but with a different field direction along [100].[9] Sakakibara et al. determined the critical point to be $(H_c, T_c) = (0.93$ T, $0.36$ K$)$ and the phase boundary as shown in Fig.10.[7] They found that $H_c(T)$ varies linearly with a positive slope of $dH_c/dT = 0.08$ T/K..

Here we will try to understand the magnitude of $H_c(T)$ within the framework of simple thermodynamics. Generally, a phase transition is described by using Gibbs' free energy $G$, which is given by $G = U - TS + PV$. Since in a magnetic system $P$ and $V$ should be replaced by $H$ and $-M$, respectively, $G = U - TS - MH$ in the present case. The internal energy $U$ arises from magnetic interactions between spins. It is known in $Dy_2Ti_2O_7$ that there are two major interactions between nearest-neighbor spins; exchange and dipole-dipole interactions.[2] The former is antiferromagnetic with $J_{nn} = -1.24k_B$, while the latter is ferromagnetic with $D_{nn} = 2.35k_B$. Then, an effective nearest-neighbor interaction $J_{eff} = J_{nn} + D_{nn} \sim 1.1k_B$ which is ferromagnetic. Taking account of only this $J_{eff}$, the $U$ for the two-in, two-out state (Fig. 1b) becomes $-4J_{eff} + 2J_{eff} = -2J_{eff}$ per tetrahedron, which is $-J_{eff}$ per Dy atom. On the other hand, the $U$ is zero for the fully saturated state (Fig. 1c) with an equal number of ferromagnetic and antiferromagnetic bonds. The $S$ is 0.65 JK$^{-1}$mol$^{-1}$ and zero for the kagomé ice state and the saturated state, while the $M$ per Dy atom is ideally $(10/3)\mu_B$ and $(10/2)\mu_B$, respectively.

Thus, we obtain $G$ for the kagomé state, $G_k$, and the fully saturated state, $G_s$, as follows:

$$G_k = -J_{eff} - \frac{0.65}{N_A}T - \frac{10}{3}\mu_B H,$$

$$G_s = -\frac{10}{2}\mu_B H,$$

where $N_A$ is Avogadro's number. Introducing a normalized field $H' = 10\mu_B H/J_{eff}$,

$$\frac{G_k}{J_{eff}} = -1 - 0.071T - \frac{1}{3}H',$$

$$\frac{G_s}{J_{eff}} = -\frac{1}{2}H'.$$

These two lines as a function of $H'$ cross at a certain critical field $H_c'$ as shown in Fig. 12. Apparently, the kagomé ice state is energetically favorable below $H_c'$, while the saturated state above $H_c'$. $H_c'$ is given from $G_k = G_s$ as $H_c' = 6 + 0.43T$. Therefore, we have obtained that $H_c(0) = 0.98$ T and $dH_c/dT = 0.070$ T/K. These values are in good agreement with the experimental values. The effect of the entropy term in G is to push the $G_k$ line downward, though this shift is relatively small; $G_k/J_{eff} = -0.028$ at $T = 0.4$ K.

The above discussion using free energy is identical to what one expects from the Clausius-Clapeyron equation for the first order phase transition:

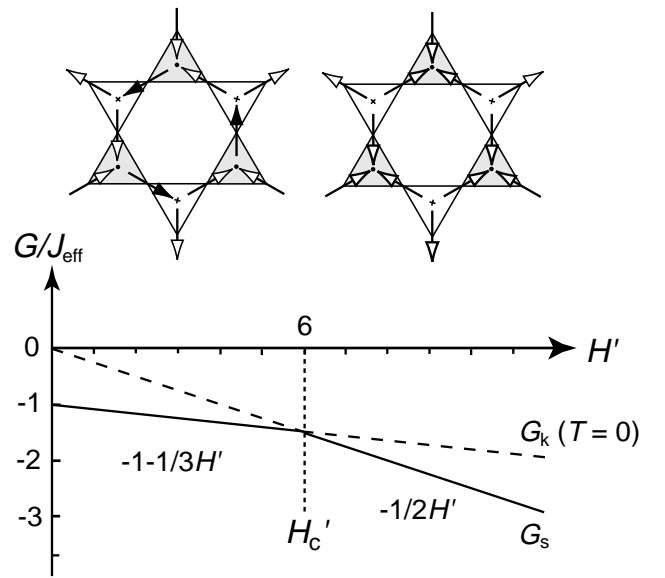

Fig. 12. Gibbs' free energy as a function of field for the ground state of the kagomé ice state $G_k$ and the fully saturated state $G_s$. $H'$ is $10\mu_B H/J_{eff}$.

$$\frac{dH_c}{dT} = -\frac{\Delta S}{\Delta M},$$

where $\Delta S$ and $\Delta M$ are jumps in $S$ and $M$ at $H_c$, respectively. As mentioned above, in the ideal case, $\Delta S = 0.65$ JK$^{-1}$mol$^{-1}$ and $\Delta M = 1.67\ \mu_B$. In our experiments, however, the $\Delta S$ must be considerably smaller than that, as shown in Fig. 7b: The $S$ already starts to decrease at $H = 0.6$ T. Unfortunately, we could not determine the $\Delta S$ in our specific heat measurements using the relaxation method. Correspondingly, magnetization measurements found that $\Delta M$ is also smaller than the ideal value, about $1\ \mu_B$.[7] Further experiments are neccesary to determine these parameters.

It would be intuitive to compare this $l$-$g$ phase transition with a typical one such as seen in H$_2$O. Figure 13a shows schematically a series of isothermal lines in a $P$-$V$ phase diagram for an ordinary $l$-$g$ transition: Below $T_c$, we have a finite jump in $V$ between the two phases. In the present case, a similar $M$-$H$ diagram is expected as a result of exchanging $P$ and $V$ by $H$ and $-M$, as shown in Fig. 13b. Here a jump in $M$ is expected below $T_c$, in good agreement with recent experiments reproduced in Fig. 14.[7] It is reasonably understood from this comparison that the $H_c(T)$ increases with temperature, as found in the magnetization measurements down to 50 mK.[7] Moreover, there appear two "plateaus" in $M$ above and below $H_c(T)$, which is simply due to small compressibility in each single-phase region. The relatively larger compressibility for the gas phase results in the larger slope in $M$ below $H_c(T)$, as observed experimentally. Therefore, from this analogy, we conclude that the kagomé ice state is a "gas" phase and the fully saturated state is a "liquid" phase. This is consistent with the fact that the former has larger entropy than the latter.

The present phase transition is quite interesting in the sense that it is probably the first example of a $l$-$g$ phase transition in localized spin systems,[7] where in most cases a phase transition takes place from a "solid" with long-range order. The present pyrochlore compound will provide us with a suitable playground to study it experimentally. It is to be marked that most features of the phase transition can be understood within the spin ice model assuming only nearest-neighbor interactions. Deviations due to long-range dipolar interactions, which have been discussed a lot,[2] may not be crucial for the present case. Nevertheless, we would naively expect an additional feature which is not seen in the ordinary case, because the residual zero-point entropy must play a crucial role in the transition. For example, the metamagnetic change in the $MH$ curves near $H_c$ below $T_c$, as shown in the inset to Fig. 14, exhibits a step-like increase at low field, while it is round at high field. This is also the case even at very low temperature down to 30 mK, and may not be due to thermal effect.[7] It may reflect some unique property for the phase transition.

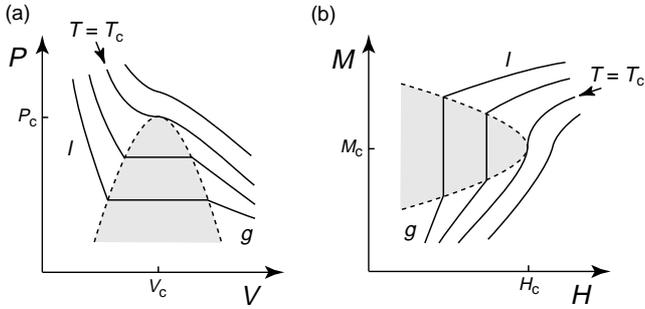

Fig. 13. Comparison between an ordinary liquid-gas phase transition in the pressure-volume diagram (a) and the present phase transition in Dy$_2$Ti$_2$O$_7$ in the magnetization-field diagram (b). A series of isothermal curves are shown in each case.

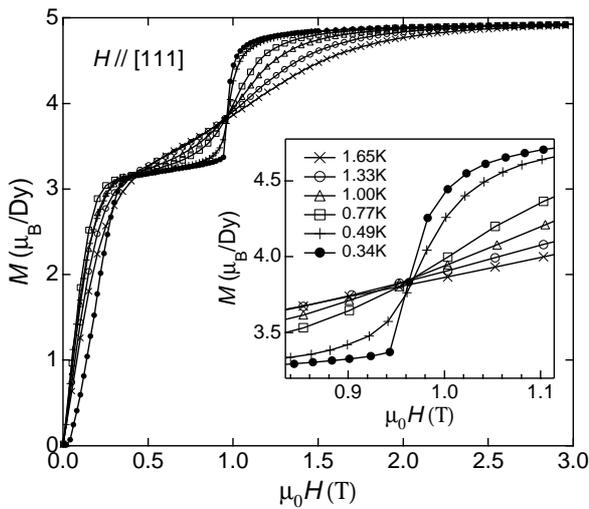

Fig. 14. Magnetization versus field curves measured at $H$ // [111] and various temperatures down to 0.34 K. As shown in the enlarged figure in the inset, the curve substantially changes from a S-shape at $T \geq 0.49$ K to a step-like shape at $T = 0.34$ K below $T_c \sim 0.36$ K.[7]

## §5. Conclusions

We have studied a macroscopically degenerate ground state called kagomé ice found in the pyrochlore oxide Dy$_2$Ti$_2$O$_7$ under magnetic field parallel to the [111] direction, by means of specific heat measurements. The first-order phase transition from the kagomé ice with residual entropy of 0.65 JK$^{-1}$mol$^{-1}$ to a fully saturated state with no residual entropy has been discussed in terms of a liquid-gas phase transition.

## Acknowledgments

We are grateful to M. Ogata and K. Ueda for their enlightening discussions. This research was supported by a Grant-in-Aid for Scientific Research on Priority Areas (A) and a Grant-in-Aid for Creative Scientific Research provided by the Ministry of Education, Culture, Sports, Science and Technology, Japan.